\RequirePackage{amsmath,amssymb}
\documentclass[runningheads]{llncs}
\usepackage[T1]{fontenc}
\usepackage[utf8]{inputenc}
\usepackage{microtype}
\usepackage{graphicx}
\usepackage{booktabs}
\usepackage{url}
\usepackage[hidelinks]{hyperref}
\usepackage{xurl}
\usepackage{enumitem}
\usepackage{xcolor}
\usepackage{pifont}
\usepackage{marvosym}
\usepackage{orcidlink}

\usepackage{algorithm}
\usepackage{algpseudocode}

\algrenewcommand\algorithmiccomment[1]{\hfill\(\triangleright\)~#1}

\makeatletter
\renewcommand\@fnsymbol[1]{\ensuremath{\ifcase#1\or\dagger\or\ddagger\or\star\or{\star\star}\else\@ctrerr\fi}}
\makeatother

\begin{document}
\title{LNTest: A Testbed for Evaluating Bitcoin Lightning Network-Based Botnets}
\titlerunning{LNTest}
\author{Thomas Bakaysa\inst{1}\thanks{These authors contributed equally to this work.} \and
Ahmet Kurt\inst{1}$^{\dagger}$\,\raisebox{0.45ex}{\small(\Letter)}\,\orcidlink{0000-0002-7175-1739} \and
Abdul-Salem Beibitkhan\inst{2}\,\orcidlink{0009-0002-1799-2213} \and \\
Jesus Maria Romo Diaz de Leon\inst{1} \and
Tag Kalat\inst{1} \and
Joshua Kramer\inst{1} \and \\
Estela Rodriguez\inst{1} \and
Abraham Watkins\inst{1} \and
Abdullah Aydeger\inst{3}\,\orcidlink{0000-0003-3333-1941}}
\authorrunning{T. Bakaysa and A. Kurt et al.}
\institute{East Texas A\&M University, RELLIS Campus, Bryan, TX 77807, USA\\
\email{tbakaysa@leomail.tamuc.edu, ahmet.kurt@etamu.edu}\\
\email{\{jromodiazde,tkalat,jkramer2\}@leomail.tamuc.edu}\\
\email{\{erodriguez80,awatkins12\}@leomail.tamuc.edu}
\and
North American University, Stafford, TX 77477, USA\\
\email{abyeibitkhan@na.edu}
\and
Florida Institute of Technology, Melbourne, FL 32901, USA\\
\email{aaydeger@fit.edu}}
\maketitle
\begin{abstract}
Bitcoin's Lightning Network (LN) can be exploited as a covert, low-cost command-and-control (C\&C) channel for botnets, as demonstrated by the LNBot and D-LNBot designs. However, both remain proof-of-concept prototypes evaluated only through simulation, leaving key questions about real-world topology formation, propagation complexity, and resilience to takedowns unanswered. We present LNTest, the first reusable testbed for LN-based botnets, built from Core Lightning nodes containerized with Docker over a shared Bitcoin Core \textit{regtest} chain. LNTest supports three overlay topology modes (a deterministic chain, autonomous peer discovery, and user-supplied graphs), enabling controlled experiments across different botnet structures. Using LNTest, we report three main findings. First, D-LNBot's autonomous formation protocol does not produce the uniform chain from its design; instead, it creates a \emph{clustered chain} in which cliques are linked by bridge nodes whose removal fragments the network. Second, command propagation scales linearly with botnet size ($\Theta(n)$), not the $O(m \log n)$ previously claimed, and gains nothing from higher neighbor connectivity. Third, the overlay topology determines the effectiveness of takedown strategies: uniform-degree chains resist targeted removal but fragment under random failure, scale-free topologies show the opposite pattern, and the autonomous clustered chain is fragile under both, making it the most vulnerable of the three. LNTest is released as open source, with a script that reproduces all our experiments, to support reproducible research on LN-based botnet defenses.

\end{abstract}

\section{Introduction}
\label{sec:introduction}

Botnets remain a persistent and versatile threat on today's Internet. Modern botnets power large-scale distributed denial-of-service (DDoS) attacks, credential stuffing, spam, ransomware, and crypto-mining campaigns. Recent surveys show that botnets are becoming harder to detect and dismantle, partly because their command-and-control (C\&C) infrastructures are shifting from simple centralized servers to distributed overlays with no single point of failure~\cite{botnetsurvey2021}. As a result, law enforcement and defenders often struggle to keep up, even when individual infections are well understood~\cite{botnettakedowns}.

One emerging trend is to use public blockchains as C\&C channels that are difficult to censor or take down. Researchers have proposed schemes that hide commands inside Bitcoin~\cite{bitcoin} transactions or Ethereum smart contracts, including ZombieCoin~\cite{zombiecoin}, ChainChannels~\cite{chainchannels}, BlockchainBot~\cite{blockchainbot}, and SCBot~\cite{SCBot}. Real-world malware has followed suit: the Pony malware hid its C\&C server addresses inside the Bitcoin blockchain~\cite{ponymalware}, and Tsundere stored its C\&C server address in an Ethereum contract~\cite{tsundere}. These examples show that blockchain-based C\&C is no longer hypothetical.

What makes blockchains attractive for this purpose is that they offer pseudonymous accounts, are replicated across thousands of nodes worldwide, and are extremely hard to tamper with~\cite{alexopoulos2019assessing,attackingBitcoin}. Prior work has analyzed these threats and proposed countermeasures~\cite{alexopoulos2019assessing,blockchainbot}, but most efforts stay at the level of analysis or small prototypes and do not give defenders a reusable environment to test and stress botnet designs under controlled conditions.

Bitcoin's Lightning Network (LN), a payment-channel network (PCN) built on top of Bitcoin, makes the problem worse. LN enables near-instant, low-cost payments that can carry hidden data, so an attacker can send botnet commands disguised as ordinary payments, cheaply, quickly, and almost invisibly. Two recent works, LNBot~\cite{lnbot} and D-LNBot~\cite{dlnbot}, have shown that this is feasible. However, both are proof-of-concept prototypes that leave important questions open. D-LNBot claims that commands propagate in $O(m \log n)$ time but tests this only in simulation, never on real Lightning nodes. Its peer discovery protocol is presented as producing a uniform chain topology, but its actual behavior under realistic conditions has not been tested. Most importantly, neither work studies how the choice of overlay topology affects resilience to partial takedowns, the question that matters most from a defender's perspective.

These gaps cannot be filled by simulations alone; they require running real Lightning software in a setting where the overlay topology, botnet size, neighbor connectivity, injection point, and takedown strategy are all controlled variables. Existing botnet testbeds such as DDoSim~\cite{DDoSim} and DDoShield-IoT~\cite{DDOSHIELD} focus on traditional IoT and DDoS botnets and do not support blockchain- or payment-channel-based C\&C. To the best of our knowledge, no testbed exists that (i)~runs on unmodified, production-grade Bitcoin and Lightning software, (ii)~scales to hundreds of C\&C servers, and (iii)~lets researchers compare different overlay topologies and takedown strategies on real nodes. To fill this gap, we introduce LNTest.

LNTest is a self-contained LN-based botnet testbed built from Bitcoin Core and Core Lightning (CLN) nodes running on Bitcoin \textit{regtest} (a private, local blockchain) and containerized with Docker. It supports three overlay topology modes: a deterministic chain matching the D-LNBot design, autonomous peer discovery following D-LNBot's formation protocol~\cite{dlnbot} under realistic deployment conditions, and user-supplied graphs that let researchers test any topology they choose (e.g., scale-free networks modeled after the real Lightning Network). Using LNTest, we conduct the first experimental study of LN-based botnet topologies on real Lightning nodes. Our \textbf{contributions} are:

\begin{itemize}[leftmargin=*]
  \item[\ding{182}] We design and implement LNTest, the first testbed for LN-based botnets. Its three topology modes let researchers directly compare how overlay structure affects command speed and resistance to takedowns.

  \item[\ding{183}] We run D-LNBot's peer discovery protocol on real Lightning nodes for the first time using LNTest's autonomous mode. We find that it does \emph{not} produce the uniform chain shown in the original paper~\cite{dlnbot}. Instead, it creates a \emph{clustered chain}: groups of tightly connected nodes linked by a small number of bridge nodes. This topology delivers commands just as fast as the ideal chain, but it is far more fragile: removing the bridge nodes splits the botnet into disconnected pieces.

  \item[\ding{184}] We measure command propagation delay across up to 500 C\&C nodes and show that propagation time grows linearly with botnet size ($\Theta(n)$), not the $O(m \log n)$ that D-LNBot claimed~\cite{dlnbot}, and is independent of the neighbor count~$m$. We also run takedown experiments (both random and targeted) across three topologies (D-LNBot chain, Barab\'{a}si--Albert (BA) scale-free, and autonomous clustered chain) and show that it is the topology, not just the size of the botnet, that determines whether a takedown fragments the network or leaves it intact.

  \item[\ding{185}] We release LNTest's source code and documentation at \url{https://github.com/ThomasBakaysaJr/LNTest} to support reproducible research and future work on detecting and defending against Lightning-based botnets.
\end{itemize}

The rest of this paper is organized as follows. Section~\ref{sec:background} provides background on LN and its use as a C\&C channel. Section~\ref{sec:relatedwork} reviews related work. Section~\ref{sec:testbed} describes the design of LNTest. Section~\ref{sec:experiments} presents our experimental results. Section~\ref{sec:discussion} discusses implications, ethical considerations, and limitations, and Section~\ref{sec:conclusion} concludes.

\section{Background}
\label{sec:background}
This section provides the background on Bitcoin's Lightning Network needed to understand LNTest's design and experiments.

\subsection{Payment Channels and the Lightning Network}
A payment channel lets two parties send money back and forth without writing every transaction to the blockchain. The two parties lock funds into a shared on-chain output and then exchange signed updates off-chain; only the final balance is settled on-chain when the channel is closed. This makes individual payments fast (seconds instead of minutes) and cheap (no mining fees per payment).

The Lightning Network (LN)~\cite{lightning} connects many such channels into a network. If Alice has a channel with Bob and Bob has a channel with Carol, Alice can pay Carol by routing the payment through Bob. The payment is secured at each hop by a hashed time-locked contract (HTLC), which ensures that either the full payment goes through or nobody loses funds. To protect privacy, LN wraps each payment in layers of encryption (onion routing) so that intermediate nodes only know their immediate predecessor and successor, not the full path.

Two LN mechanisms are especially relevant to this work. First, \emph{gossip}: LN nodes announce their channels to the network, and other nodes store these announcements in a local channel graph. This is how nodes learn about the network's topology and discover potential peers. In LNTest, the autonomous formation mode relies on gossip to let C\&C servers find each other (Section~\ref{sec:topology-control}). Second, \emph{keysend}: a way to send a payment to a node without first requesting an invoice from it.\footnote{Keysend specification: \url{https://github.com/lightning/blips/blob/master/blip-0003.md}} Keysend payments can carry custom data in type-length-value (TLV) fields that are visible only to the final recipient, not to intermediate routing nodes. All three botnet designs we discuss use keysend to deliver commands (Section~\ref{sec:command-embedding}).

\subsection{LN-Based Botnet Designs}
Two prior works have proposed using LN as a botnet C\&C channel.

\textbf{LNBot}~\cite{lnbot} uses a centralized one-to-many model: the botmaster sends each command to every C\&C server individually as a series of LN keysend payments, one per character, whose amounts encode the character via ASCII or Huffman coding. This design is simple but creates a single point of failure at the botmaster.

\textbf{D-LNBot}~\cite{dlnbot} addresses this with a distributed design. Instead of the botmaster reaching each C\&C server individually, C\&C servers form a chain-like overlay among themselves. Each node $\mathrm{CC}_i$ connects to its $m$ nearest predecessors ($\mathrm{CC}_{i-1}$ through $\mathrm{CC}_{i-m}$), creating overlapping neighborhoods. To build this chain without a central coordinator, D-LNBot introduces a formation protocol: newly infected machines connect to an \emph{innocent node} (a publicly-reachable LN node) and discover other C\&C servers by scanning the Lightning Network for channels whose capacity matches a pre-agreed rule. A node closes its innocent-node channel after $m$ newer C\&C servers have joined the network, so it is no longer discoverable. D-LNBot claims that commands propagate through this chain in $O(m \log n)$ time, based on the assumption that the topology resembles an $m$-ary tree. However, both the topology and the propagation time have only been evaluated in simulation, not on real Lightning nodes.

\subsection{Command Embedding via Keysend}
\label{sec:command-embedding}
LNBot, D-LNBot, and LNTest all send commands as Lightning keysend payments, but they carry the command differently.

LNBot encodes one character per payment using the payment amount itself, requiring multiple payments per command. D-LNBot attaches a full command as a data blob to a single keysend payment using a Core Lightning plugin.

LNTest takes a simpler approach: it places the command text and a sequence counter directly into a custom TLV field attached to a standard keysend payment. The sequence counter allows receiving nodes to detect duplicates (since flooding causes the same command to arrive from multiple neighbors) and to track which commands they have already processed and forwarded. Because TLV fields are encrypted end-to-end by LN's onion routing, intermediate nodes along the payment path cannot read the command; only the intended C\&C server can.

\section{Related Work}
\label{sec:relatedwork}
We review prior work on botnet testbeds, LN-based botnets, other blockchain-based botnets, and network resilience analysis.

\smallskip\noindent\textbf{Botnet testbeds:}
Existing botnet testbeds target traditional IoT and DDoS botnets. DDoSim~\cite{DDoSim} emulates large-scale IoT botnets in an ns-3/Docker environment; DDoShield-IoT~\cite{DDOSHIELD} extends it with Mirai-style binaries for IDS evaluation; Gotham Testbed~\cite{gothamtestbed} provides a reproducible IoT security platform; Calvet et al.~\cite{calvet2010lab} created and took down a 3000-node in-the-lab botnet; the testbed of Kumar and Lim~\cite{detertestbed} analyzes IoT botnets in a securely contained environment; and Parthipan et al.~\cite{EEI8654} review peer-to-peer botnet emulation testbeds and propose a conceptual model for them. None considers blockchains or payment-channel networks as the C\&C substrate. LNTest fills this gap.

\smallskip\noindent\textbf{LN-based botnets:}
LNBot~\cite{lnbot} and D-LNBot~\cite{dlnbot}, the two LN-based botnets introduced in Section~\ref{sec:background}, were both evaluated only in simulation. LNTest is the first to run them on production-grade CLN nodes, revealing the clustered chain that D-LNBot's autonomous protocol actually forms, with $\Theta(n)$ propagation rather than the claimed $O(m \log n)$ (Sections~\ref{sec:scalability} and~\ref{sec:formation}).

\smallskip\noindent\textbf{Blockchain-based botnets:}
Beyond LN, a range of blockchain-based C\&C designs have been proposed: on Bitcoin, the Testnet botnet~\cite{testnetbotnet}, ZombieCoin and its successor~\cite{zombiecoin,zombiecoin2}, CoinBot~\cite{coinbot}, ChainChannels~\cite{chainchannels}, DUSTBot~\cite{DUSTBot}, and Kamenski et al.'s resilient scheme~\cite{attackingBitcoin}; on Ethereum, Botract~\cite{botract}, Unblockable Chains~\cite{unblockablechains}, Sweeny's private-chain design~\cite{sweeny}, and SCBot~\cite{SCBot}; on IOTA, BlockchainBot~\cite{blockchainbot}, OICL~\cite{OICL}, and ZombieCoin~3.0~\cite{zombiecoin3}; and on Monero~\cite{monerobotnet}. All confirm the appeal of blockchain C\&C but provide individual prototypes rather than a reusable testbed for systematic topology and resilience experiments.

\smallskip\noindent\textbf{Network resilience and LN topology:}
Our takedown experiments connect to classical results in network science. Albert et al.~\cite{albert2000error} showed that scale-free networks are robust to random failure but fragile under targeted hub removal, precisely the asymmetry we observe on BA topologies in Section~\ref{sec:takedown}. Empirical studies of the real Lightning Network underscore its exposure to such targeted attacks: Zabka et al.~\cite{ln-centrality} found it highly and increasingly centralized, with the top 5\% of nodes carrying the vast majority of payment routes, and Rohrer et al.~\cite{rohrer2019discharged} demonstrated that centrality-based removal strategies can disconnect the LN with only a small fraction of nodes removed. Our contribution is to reproduce and extend these findings in a controlled botnet testbed where topology, takedown strategy, and C\&C parameters are all independent variables.

\section{LNTest Design}
\label{sec:testbed}

LNTest is designed around three goals motivated by the limitations of existing LN-based botnet prototypes~\cite{lnbot,dlnbot}: (i)~\emph{topology flexibility}, so that researchers can study how overlay structure affects propagation and resilience; (ii)~\emph{reproducibility}, so that experiments can be repeated exactly across machines and over time; and (iii)~\emph{low-overhead monitoring}, so that the orchestrator can track every node's state in real time without perturbing the system under test.

\begin{figure}[ht]
    \centering
    \vspace{-5mm}
    \includegraphics[width=\linewidth]{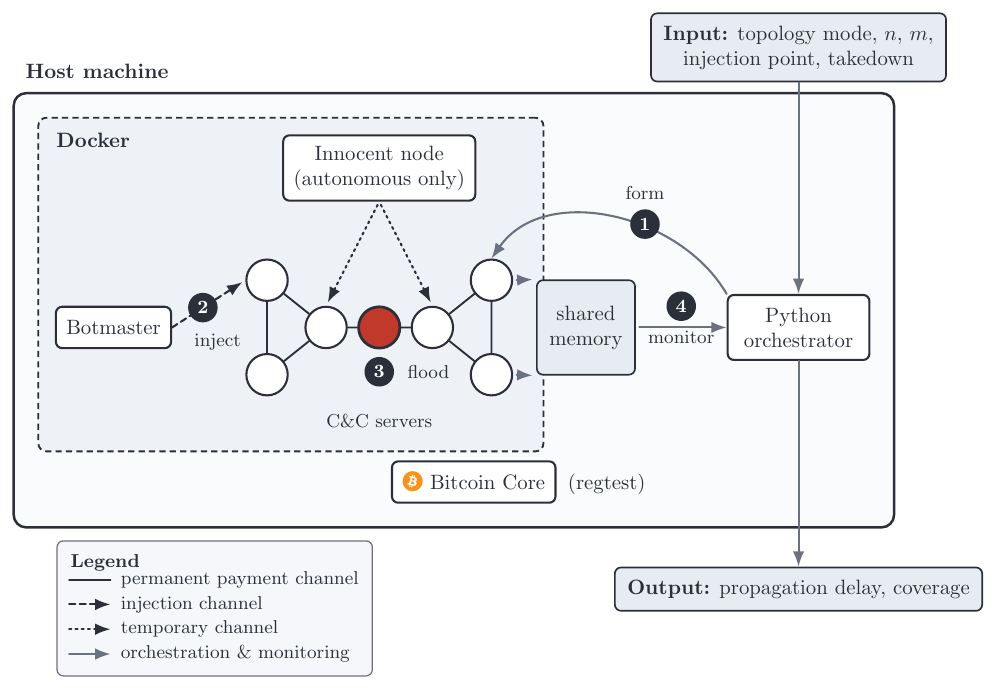}
    \vspace{-6mm}
    \caption{Overview of LNTest's architecture. The botmaster, innocent node, and C\&C servers are Core Lightning containers in Docker, while Bitcoin Core, the shared memory, and the orchestrator run on the host. On-chain links to the single \textit{regtest} Bitcoin Core are omitted for clarity. The example overlay is an autonomous clustered chain (Section~\ref{sec:formation}): cliques joined by a bridge node (red) whose removal splits it.}
    \label{fig:system_model}
\end{figure}

\subsection{Overview}
\label{sec:overview}

LNTest runs entirely on a single host machine. It uses Docker containers to isolate each Lightning node and Bitcoin \textit{regtest} to provide a private, deterministic blockchain that never touches the public Bitcoin network. Built on unmodified Bitcoin Core and Core Lightning, the system consists of four logical components, each chosen to mirror a specific role in LN-based botnet architectures:

\noindent\textbf{Bitcoin Core} runs directly on the host and provides the shared on-chain layer. All channel funding, confirmations, and closures go through this single node, giving the orchestrator full control over block production and transaction timing.

\noindent\textbf{C\&C servers} are CLN instances, each in its own Docker container, that form the botnet's overlay by opening payment channels to one another. In D-LNBot and custom modes the orchestrator opens these channels directly; in autonomous mode each container discovers peers via the innocent node and opens its own channels. Commands propagate through this overlay by flooding. The overlay topology is not fixed: LNTest supports three modes (a deterministic D-LNBot chain, autonomous peer discovery, and user-supplied graphs), letting researchers study how structure affects propagation speed and takedown resilience (Section~\ref{sec:topology-control}).

\noindent\textbf{Botmaster} is a separate CLN node that opens a channel to one or more C\&C servers to inject commands into the overlay. Because this channel can be opened to any node, the botmaster's injection point is an experimental variable: researchers can measure how entry position affects propagation delay.

\noindent\textbf{Innocent node} is a CLN node that acts as a rendezvous point during overlay formation. In autonomous mode, C\&C servers discover each other by scanning the innocent node's channel list for peers that match a shared capacity-based rule. In the other two modes, the innocent node is present but not used for discovery. Once formation is complete, the innocent node plays no further role, and the overlay operates independently.

A Python orchestrator on the host coordinates these four components and drives each experiment through the two phases described in Sections~\ref{sec:topology-control} and~\ref{sec:evaluation}. Fig.~\ref{fig:system_model} shows how they interact: the C\&C servers form the overlay through permanent payment channels, the botmaster reaches it through an injection channel, and in autonomous mode each node opens a temporary channel to the innocent node for peer discovery; the numbered badges trace the workflow from formation to monitoring. The orchestrator tracks every node's state through shared memory, and all nodes rely on Bitcoin Core for on-chain operations.

We run LNTest on Bitcoin \textit{regtest}, not Mainnet or public Testnet, because controlled, reproducible experiments need a chain we fully control. Mining blocks on demand and funding nodes locally makes timing, funding, and confirmations deterministic across runs and lets us fund the hundreds of nodes our largest tests need. Public Testnet gives none of this: even Testnet4 suffers miner-gamed block timing, frequent reorgs, and depleted faucets; Mainnet is excluded on ethical and cost grounds.\footnote{Testnet4 keeps a 20-minute minimum-difficulty rule that miners still exploit to mine empty blocks on demand; a fix remains unmerged as of June 2026 (\url{https://github.com/bitcoin/bitcoin/pull/35081}).}

\subsection{Networking}
Every Lightning node (the botmaster, the innocent node, and all C\&C servers) runs as a Core Lightning instance inside its own Docker container. All containers share the host's network namespace (\textit{-{}-network host}), and each CLN process binds to 127.0.0.1 on a unique port, so no traffic leaves the machine. Deterministic port assignment (base port plus node index) allows the orchestrator to address any node by its index without maintaining a registry. C\&C containers additionally share the host's inter-process communication (IPC) namespace (\textit{-{}-ipc=host}), which gives them write access to the shared memory buffers used for real-time status reporting (Section~\ref{sec:command-monitoring}).

Bitcoin Core runs directly on the host in \textit{regtest} mode and serves as the shared on-chain layer. Each CLN instance connects to it via JSON-RPC on localhost and polls it for new blocks. As a result, all on-chain activity, peer-to-peer Lightning traffic, and orchestration commands stay entirely within the host, and LNTest never contacts the public Bitcoin network.

\subsection{Overlay Formation}
\label{sec:topology-control}

We now describe how LNTest forms the C\&C overlay; Section~\ref{sec:evaluation} then covers how it evaluates resilience. To build the overlay, the orchestrator initializes the regtest chain (reused across sweep iterations), launches all containers, and either wires the channels itself or waits for the nodes to form them (Algorithm~\ref{alg:formation}).

LNTest offers three ways to create the overlay, each targeting a different research question.

\smallskip\noindent\textbf{D-LNBot mode:} The orchestrator builds the sequential chain described in the D-LNBot paper~\cite{dlnbot}: each node $\mathrm{CC}_i$ opens channels to its $m$ immediate predecessors ($\mathrm{CC}_{\max(1,\,i-m)}$ through $\mathrm{CC}_{i-1}$); together with the $m$ successors that later open channels to it, each interior node ends up with $2m$ channels. Because the orchestrator controls every channel, the result is deterministic and reproducible across runs.

\smallskip\noindent\textbf{Custom mode:} The orchestrator reads an arbitrary graph from a JSON file and wires exactly those channels. This lets researchers test any topology they choose (scale-free graphs, small-world networks, real LN snapshots, or hand-crafted structures) on real Lightning nodes.

\smallskip\noindent\textbf{Autonomous mode:} This mode simulates a realistic D-LNBot deployment: rather than wiring the topology from outside, each node discovers its own peers (Algorithm~\ref{alg:formation}). Containers launch one at a time with random delays (log-normal, median 30\,s, clamped to $[10, 90]$\,s), reflecting the variability of D-LNBot's malware pipeline, where a newly infected machine must download a Lightning client, sync, obtain funding, and wait for channel confirmations before it can participate. Each node advertises itself and finds peers through a shared capacity-based rule (a pre-agreed funding amount that distinguishes C\&C channels from ordinary ones); a small Core Lightning plugin caps each node at $2m$ inbound C\&C channels; this keeps a node that is briefly flooded with connection requests from ballooning into an oversized hub, and it has no other effect on the topology. Because multiple nodes scan at the same time, they discover each other mutually, so the topology that emerges is \emph{not} the clean chain from D-LNBot's paper; we characterize it experimentally in Section~\ref{sec:formation}.

\vspace{-3mm}
\begin{algorithm}[!ht]
\footnotesize
\caption{Overlay Formation}
\label{alg:formation}
\begin{algorithmic}[1]
\raggedright
\Require $n$ C\&C nodes; active neighbors per node $m$; mode $M\!\in\!\{\textsc{dlnbot},\textsc{custom},\textsc{autonomous}\}$; edge set $F$ (if $M{=}\textsc{custom}$)
\Ensure overlay $G=(V,E)$
\State initialize regtest; launch Innocent, Botmaster; $V \gets \{\mathrm{CC}_1,\dots,\mathrm{CC}_n\}$
\If{$M = \textsc{autonomous}$}
    \State \textbf{for} $i = 1\dots n$: launch and fund $\mathrm{CC}_i$; wait $\delta_i$ \Comment{$\delta_i\!\sim$ lognormal, med.\ $30$\,s}
    \State each $\mathrm{CC}_i$ runs \Call{Discover}{$i$} once online; $E \gets$ formed channels
\Else
    \State \textbf{if} $M{=}\textsc{dlnbot}$ \textbf{then} $E \gets \{(i,j):\max(1,i{-}m)\le j<i\}$ \textbf{else} $E \gets F$
    \State launch and fund $V$; open all channels in $E$
\EndIf
\State \Return $G$
\Procedure{Discover}{$i$} \Comment{runs on $\mathrm{CC}_i$}
    \State advertise: open a discovery-rule channel to Innocent
    \State plugin: reject inbound opens once $\deg_{\mathrm{in}}(\mathrm{CC}_i){=}2m$
    \State \textbf{while} $\deg_{\mathrm{in}}(\mathrm{CC}_i){<}m$: open $\le m$ channels to new advertisers in Innocent's channels
    \State close Innocent channel
\EndProcedure
\end{algorithmic}
\end{algorithm}
\vspace{-3mm}

In D-LNBot and custom modes the per-node channel manager is disabled, so the orchestrator keeps full control and opens exactly the prescribed channels; in autonomous mode each node instead runs the discovery loop of Algorithm~\ref{alg:formation} concurrently with the others. This concurrency is the crux of our formation finding: D-LNBot's formation protocol~\cite{dlnbot} is an idealized, effectively sequential procedure meant to build a uniform chain (which its complexity analysis then models as an $m$-ary tree), whereas Algorithm~\ref{alg:formation} reflects how formation actually executes, with many nodes discovering peers at once. The result is the clustered chain of Section~\ref{sec:formation}, not the uniform chain D-LNBot intends.

\subsection{Command Dissemination and Monitoring}
\label{sec:command-monitoring}

Once the topology is established, commands spread through the overlay by flooding. The botmaster injects a command at one or more C\&C servers (the injection points) as a Lightning \textit{keysend} payment, with the command text embedded in a custom TLV (type-length-value) field. When a C\&C node receives a new command, it immediately forwards the same payment to every neighbor it has a channel with. These forwards are dispatched in parallel, one thread per neighbor, rather than sequentially. Every command carries a sequence counter, and each node tracks which counters it has already processed. If a node receives the same command from a second neighbor, it simply ignores the duplicate. This ensures that every reachable node receives each command and forwards it only once, despite the redundant copies that flooding delivers.

Reception is event-driven: a node reacts the moment a keysend arrives rather than polling on a fixed interval, so it forwards immediately. Each keysend arrives as a paid invoice, from which the command is extracted.

\smallskip\noindent\textbf{Propagation complexity:} The chain's structure explains why delay is linear. With mid-chain injection, the farthest node is about $n/2$ positions away, and each flooding hop advances the command about $m$ positions, so coverage takes $H \approx n/(2m)$ hops. The per-hop time $\tau$ is proportional to $m$: the relaying node settles a keysend to each of its up to $2m$ neighbors, which a single Core Lightning daemon does largely one at a time. The two factors cancel, $T = H\tau \approx (n/(2m))\cdot m = \Theta(n)$: a wider neighborhood shortens the path but makes each hop proportionally heavier. The bound is tight, not just an upper limit: the wavefront must cross all $H$ hops, each costing $\propto m$, so linear growth is a lower bound as well, giving $\Theta(n)$ rather than merely $O(n)$. Delay is therefore linear in $n$ and flat in $m$, not the logarithmic $O(m \log n)$ of D-LNBot's $m$-ary-tree model.

\vspace{-3mm}
\begin{algorithm}[!ht]
\footnotesize
\caption{Resilience Evaluation}
\label{alg:evaluation}
\begin{algorithmic}[1]
\raggedright

\Require overlay $G=(V,E)$; $k$ commands per level; removal levels $P$ ($\{0\%\}$ if none); strategy $S\!\in\!\{\textsc{random},\textsc{targeted}\}$
\Ensure delay $t_{p,j}$ and coverage $c_{p,j}$ for every level $p$, command $j$
\State $R \gets$ rank $V$ on intact $G$ \Comment{by degree ($\textsc{targeted}$) or fixed-seed shuffle ($\textsc{random}$)}
\For{$p \in P$}
    \State remove $R[1\dots\lfloor pn\rfloor]$ from $G$ \Comment{cumulative}
    \State $v \gets$ highest-degree node in $G$'s largest surviving component
    \For{$j = 1 \dots k$}
        \State Botmaster keysends command $j$ at $v$; each node floods it onward
        \State record $t_{p,j}, c_{p,j}$ \Comment{$60$\,s stall before full coverage $\Rightarrow$ partition}
    \EndFor
\EndFor
\State \Return $\{(t_{p,j}, c_{p,j})\}$

\end{algorithmic}
\end{algorithm}
\vspace{-3mm}

To give the orchestrator a real-time view of the network without the overhead of reading log files, LNTest uses POSIX shared memory (SHM). Before launching the containers, the orchestrator allocates a small status buffer for each C\&C node. As a node creates channels, joins the overlay, and forwards commands, it writes its current state into this buffer. The orchestrator simply reads the buffers to track how far each command has propagated, which nodes are online, and when coverage has stalled, all with minimal latency.

\subsection{Resilience Evaluation}
\label{sec:evaluation}

LNTest measures resilience by simulating a takedown, which Algorithm~\ref{alg:evaluation} formalizes. It shuts down some of the C\&C nodes and checks whether commands still reach the rest. Nodes are removed either at random or in a targeted way that takes the highest-degree nodes first. The takedown grows step by step, and each step adds to the nodes already removed. This makes it one steadily worsening attack rather than a fresh cut each time. A takedown can split the overlay into several disconnected components, so after each step the botmaster injects from the highest-degree node in the largest surviving component to maximize its reach. A command whose coverage stalls for 60~seconds before reaching every survivor counts as a partition.

\section{Experiment Results}
\label{sec:experiments}

We now present our experimental results. Our primary metric is \textit{propagation delay}: the time from when the botmaster sends a command until the last surviving C\&C node receives it. For every data point we issue the command 10 times and report the mean propagation delay; the repeated sends average out per-command jitter. If the network partitions, the orchestrator detects the stall and stops early. We also report \textit{coverage}: the fraction of surviving nodes that received the command, which matters in takedown experiments where the network may be partitioned.

All experiments ran on a single workstation: a Supermicro tower with a 16-core AMD Ryzen Threadripper PRO 3955WX CPU and 64\,GB DDR4 3200\,MT/s RAM, running Ubuntu~26.04 on a 1\,TB Toshiba KXG60ZNV1T02 NVMe SSD (PCIe~Gen3 $\times4$). Fast storage proved essential: an initial install on the machine's 2\,TB 7200\,RPM SATA hard drive was unusably slow because the many co-located Lightning nodes generate heavy concurrent disk I/O, so we moved the system to the NVMe SSD. The software stack is Bitcoin Core\footnote{\url{https://github.com/bitcoin/bitcoin}} version~31.0, Core Lightning~\cite{cln} version~26.06rc2, and Docker~29.5.2.

\subsection{Scalability}
\label{sec:scalability}

\begin{figure}[ht]
  \centering
  \includegraphics[width=0.9\linewidth]{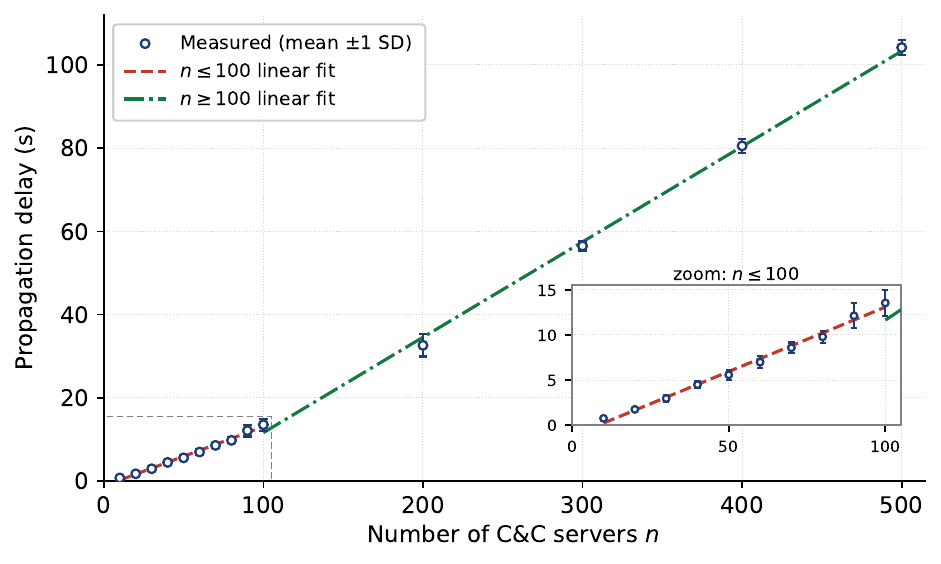}
  \vspace{-4mm}
  \caption{Propagation delay vs.\ number of C\&C servers ($m = 4$, D-LNBot chain, mid-chain injection). Markers show the mean of 10 commands with $\pm 1$ standard deviation error bars; coverage is 100\% at every size. The two regimes are fit separately: $t = 0.143n - 1.20$ ($R^2 = 0.993$) for $n \leq 100$ and $t = 0.229n - 11.28$ ($R^2 = 0.998$) for $n \geq 100$. The inset zooms into the $n \leq 100$ regime.}
  \label{fig:scalability}
  \vspace{-3mm}
\end{figure}

We first ask how propagation delay grows with botnet size. Using D-LNBot mode with $m = 4$ active neighbors and injecting each command from the middle of the chain, we varied the number of C\&C servers from 10 to 100 in steps of~10, and then from 100 to 500 in steps of~100 to probe behavior well beyond proof-of-concept scale. Fig.~\ref{fig:scalability} shows the results.

Through $n = 100$, propagation delay grows linearly with the number of nodes: a fit gives $t = 0.143n - 1.20$ ($R^2 = 0.993$), so each additional C\&C server adds about 0.14\,s. At $n = 100$ the mean delay is 13.5\,s, and coverage is 100\% in every run, with no command ever lost. The trend continues well past proof-of-concept scale. At $n = 500$ commands still reach every node in 104\,s on average, again at 100\% coverage.

This linear scaling confirms the $\Theta(n)$ analysis of Section~\ref{sec:command-monitoring} and contradicts the $O(m \log n)$ that D-LNBot~\cite{dlnbot} predicted by treating the chain as an $m$-ary tree. The real overlay is a chain of overlapping neighborhoods, not a tree, so the command wavefront advances only a bounded number of positions per hop and the delay grows linearly with~$n$, governed by botnet size rather than neighbor count, far slower than the logarithmic growth D-LNBot predicted.

The linear fit is near-perfect through $n = 100$, the range relevant to existing proof-of-concept designs; beyond it a second, steeper linear regime appears, with the per-node cost rising from 0.14\,s to 0.23\,s (Fig.~\ref{fig:scalability}). This drift is not a path-length effect, since the number of hops from the injection point to the farthest node stays at exactly $\lceil n/(2m) \rceil$ (we verified this on the reconstructed overlays), nor a host-resource limit, since even at $n = 500$ the workstation uses only about half of its 16 CPU cores on average (Section~\ref{sec:hw-resources}). Instead, each forwarding hop settles more slowly as the overlay grows. The cost is not route computation, which stays negligible (a few milliseconds), but the per-keysend HTLC settlement, which lengthens as each node services a larger, busier channel graph. The conclusion is unchanged: propagation is linear in~$n$, so a larger botnet is strictly slower, never faster.

\subsection{Autonomous Formation Topology}
\label{sec:formation}

A key question is what topology D-LNBot's formation protocol actually produces when run on real Lightning nodes under realistic conditions. As described in Section~\ref{sec:topology-control}, the autonomous mode launches containers with staggered delays (log-normal, median 30\,s) and lets each node discover peers by scanning the innocent node's channels. We ran this mode at two scales: $n = 20$ and $n = 50$, both with $m = 4$. Each command was injected from $\mathrm{CC}_{\lceil n/2 \rceil}$, as in the scalability runs.

\begin{figure}[ht]
  \centering
  \vspace{-2mm}
  \includegraphics[width=\linewidth]{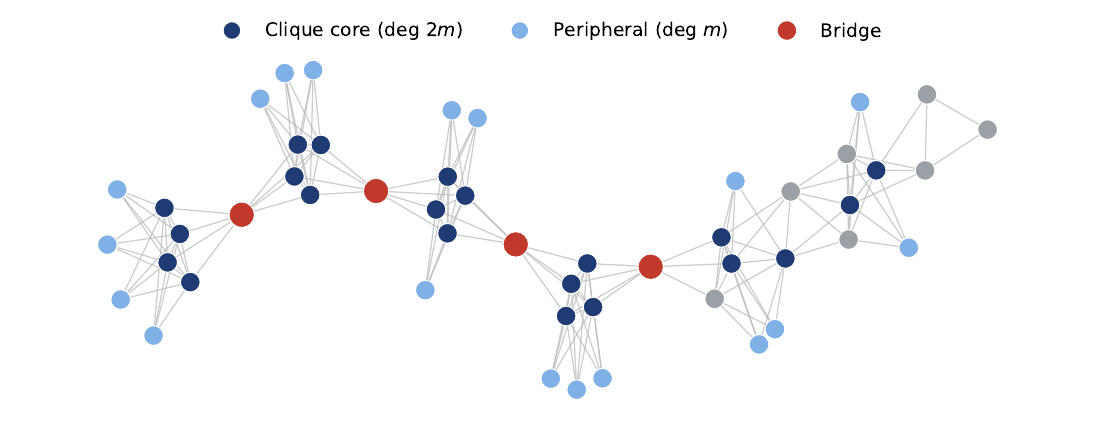}
  \vspace{-8mm}
  \caption{A representative overlay produced by D-LNBot's autonomous formation protocol at $n = 50$, $m = 4$: a clustered chain of cliques (cores at degree~$2m = 8$, dark blue; peripheral nodes at degree~$m = 4$, light blue) joined by bridge nodes (red) whose removal splits the overlay; intermediate-degree nodes are gray. Because formation is nondeterministic, the bridge count varies across runs.}
  \label{fig:formation_topology}
  \vspace{-2mm}
\end{figure}

\smallskip\noindent\textbf{Results:} At both scales the formation protocol produced a connected overlay that delivered every command to all nodes (100\% coverage; mean delay 2.0\,s at $n = 20$, 5.45\,s at $n = 50$, averaged over five formations). But the topology was not the uniform chain of D-LNBot's paper. Instead we observed a \emph{clustered chain} (Fig.~\ref{fig:formation_topology}): a sequence of cliques of roughly $m$ nodes at degree~$2m$, each ringed by peripheral nodes at degree~$m$ and joined to its neighbors by bridge nodes whose removal would split the overlay. The degree distribution is roughly bimodal, peaking at the clique-core degree ($2m = 8$) and the peripheral degree ($m = 4$). Because the overlay forms nondeterministically, its structure varies across runs at $n = 50$ (two to four bridge nodes in our five formations); at $n = 20$, by contrast, all five formations produced the \emph{identical} clustered chain with two bridges (CC9 and CC17), so this run-to-run variation emerged only at $n = 50$.

\smallskip\noindent\textbf{Why cliques form instead of a chain:} The key insight is that D-LNBot's formation protocol includes continuous peer monitoring: each node keeps scanning for new peers as long as it is active. When multiple nodes have overlapping active periods (inevitable with staggered arrivals), they discover each other mutually and all open channels to one another, forming a clique. Because each node keeps accepting peers until $m$ have connected to it, and a new node arrives roughly every 30\,s (the median inter-arrival), about $m$ nodes are forming at the same time, so they interconnect into a clique of size $\approx m$. Once its members reach $m$ inbound connections they disconnect from the innocent node, and the next arrival starts a fresh cluster. This process repeats, producing the observed chain of cliques.

\smallskip\noindent\textbf{Comparison with the ideal D-LNBot chain:} Table~\ref{tab:formation_comparison} compares the two topologies at $n = 50$.

\begin{table}[ht]
  \centering
  \caption{Ideal D-LNBot chain vs.\ the autonomous clustered chain ($n = 50$, $m = 4$), both injected from $\mathrm{CC}_{\lceil n/2 \rceil}$ (CC25). Clustered-chain values are means over five formations (Fig.~\ref{fig:formation_topology} shows a representative one); its bridge count varies across runs.}
  \label{tab:formation_comparison}
  \setlength{\tabcolsep}{12pt}
  \begin{tabular}{@{}l c c@{}}
    \toprule
    \textbf{Property} & \textbf{D-LNBot chain} & \textbf{Clustered chain} \\
    \midrule
    Avg.\ degree         & 7.6  & 6.2 \\
    Diameter             & 13   & 12  \\
    Bridge nodes         & 0    & 2--4 \\
    Mean delay (10 msgs) & 5.58\,s & 5.45\,s \\
    Coverage             & 100\% & 100\% \\
    \bottomrule
  \end{tabular}
\end{table}

The two topologies have comparable diameter and propagation speed. But they differ fundamentally in vulnerability: the ideal chain has no bridge nodes (every interior node has $2m$ connections, giving many redundant paths), whereas the clustered chain always contains bridge nodes, each a single point of failure whose removal fragments the overlay (in the representative run of Fig.~\ref{fig:formation_topology}, removing the four bridges splits it into five components). We evaluate this vulnerability quantitatively in the takedown experiments below.

\subsection{Resilience to Takedowns}
\label{sec:takedown}

\begin{figure}[ht]
  \centering
  \includegraphics[width=\linewidth]{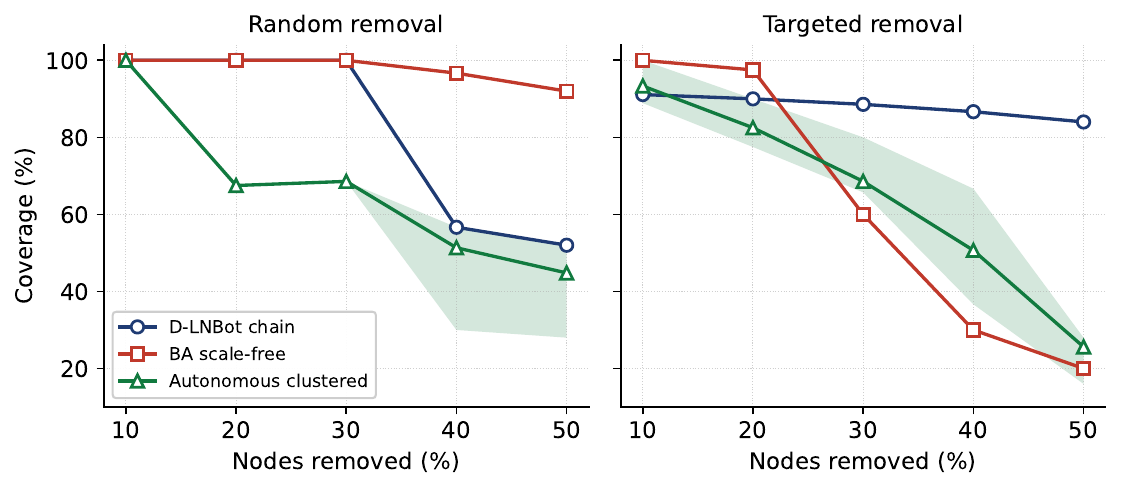}
  \vspace{-7mm}
  \caption{Coverage vs.\ fraction of nodes removed for three topologies under random and targeted takedown ($n = 50$, $m = 4$), with each command injected from the highest-degree node in the largest surviving component. A drop below 100\% means the network has partitioned. For the autonomous topology, the only nondeterministic one, the line is the mean over five runs and the shaded band their min--max range.}
  \label{fig:takedown_coverage}
  \vspace{-2mm}
\end{figure}

We now ask the central question for defenders: how much of the botnet must be taken down to stop commands from reaching all nodes, and does the answer depend on topology? We test three topologies (the D-LNBot chain, a BA scale-free graph, and the autonomous clustered chain from Section~\ref{sec:formation}) under two strategies, random removal and targeted removal of the highest-degree nodes, at five removal percentages (10\%--50\%). All runs use $n = 50$ and $m = 4$ with cumulative removal. After each removal step the botmaster re-injects from the highest-degree node in the largest surviving component (Section~\ref{sec:evaluation}), so its reach is never wasted on a fragment it cannot leave. The D-LNBot chain and the BA graph are fixed, and the random removal order is seeded, so a takedown on them is fully reproducible and we run each once. Only the autonomous topology is nondeterministic, since a fresh graph forms on every run, so we repeat it five times and plot the mean coverage with shaded min--max bands (Fig.~\ref{fig:takedown_coverage}). The specific graph varies across the five runs, but the structural patterns hold.

\smallskip\noindent\textbf{D-LNBot chain:} Under random removal, the chain stays intact through 30\% (100\% coverage, delays around 5\,s) and partitions at 40\%, where coverage falls to 57\%. Random removal scatters the missing nodes along the chain, creating several disconnected fragments, and the botmaster reaches only the nodes in its own fragment.

Targeted removal partitions the chain at the very first level (10\%), yet the resulting coverage (91\%) is \emph{higher} than random removal at 40\% (57\%), and it stays high as more nodes fall (84\% even at 50\%). This counterintuitive result has a structural explanation. In the D-LNBot chain, 42 of the 50 nodes share the same degree ($2m = 8$); only the four nodes at each end have lower degrees. Because these interior nodes are all tied at degree~8, the targeted strategy takes them in index order, always removing a consecutive block (CC5--CC9 at 10\%, CC5--CC14 at 20\%, and so on) that makes a single clean cut and orphans only the four lowest-indexed nodes (CC1--CC4). The botmaster sits in the larger fragment and reaches almost every surviving node there. Degree-based targeting thus provides \emph{no useful signal} on the chain, where almost every node has the same degree: ranking C\&C nodes by degree reveals nothing about which removals would actually fragment the overlay.

\smallskip\noindent\textbf{BA scale-free:} The BA graph tells the opposite story. Random removal barely dents it: coverage holds at 100\% through 30\% and slips only to 97\% at 40\% and 92\% at 50\%, when a few low-degree nodes lose all their links. This is because random removal is unlikely to hit the small number of high-degree hubs that hold the graph together.

Targeted removal, by contrast, is devastating. Coverage is still 100\% at 10\%, but then collapses as the hubs fall: 98\% at 20\%, 60\% at 30\%, 30\% at 40\%, and just 20\% at 50\%. The most-connected hubs are removed first; the graph absorbs the loss of the first few, but once enough are gone the periphery they held together loses its links and splinters off, so the giant component shrinks with every further removal.

\smallskip\noindent\textbf{Autonomous clustered chain:} The autonomous topology is the only one fragile under \emph{both} strategies; the values here are the mean of five fresh formations. Random removal partitions it almost immediately, coverage falling from 100\% at 10\% to about 68\% at 20\% and drifting down to roughly 45\% at 50\%, because the clustered chain has far fewer redundant paths than the D-LNBot chain or the BA graph, so even scattered removals tend to hit a bridge. Targeted removal bites early too (93\% at 10\%) and then declines steadily, from 83\% at 20\% to 26\% at 50\%: removing the highest-degree nodes strips out the dense clique cores and, as it proceeds, the high-degree bridges that join the clusters (Section~\ref{sec:formation}), progressively splitting the overlay. Either way the overlay fragments well before half its nodes are gone, making it the most vulnerable of the three: every other topology withstands at least one strategy, whereas the clustered chain fragments under both.

\smallskip\noindent\textbf{Key insight:} topology determines resilience, not just size. All three topologies have the same size and the same $m$, yet they respond to takedowns in completely different ways: the uniform-degree chain resists targeted removal but fragments under heavy random failure, the BA graph shows the mirror image, and the clustered chain is fragile to both.

\subsection{Effect of Active Neighbor Count}
\label{sec:active-nodes}

In D-LNBot's chain topology, each C\&C server opens channels to its $m$ nearest predecessors. We vary $m$ from 2 to 6 at three scales ($n = 50$, $n = 100$, and $n = 200$) to measure how the neighbor count affects propagation delay. Table~\ref{tab:active-nodes} summarizes the results.

\begin{table}[ht]
  \centering
  \caption{Mean propagation delay (seconds) vs.\ active neighbor count~$m$, with each command injected from the middle of the chain ($\mathrm{CC}_{\lceil n/2 \rceil}$).}
  \label{tab:active-nodes}
  \setlength{\tabcolsep}{12pt}
  \begin{tabular}{@{}c c c c@{}}
    \toprule
    \textbf{$m$} & \textbf{$n = 50$} & \textbf{$n = 100$} & \textbf{$n = 200$} \\
    \midrule
    2 & 5.60 & 11.79 & 26.20 \\
    3 & 5.45 & 12.67 & 29.74 \\
    4 & 5.74 & 13.46 & 32.41 \\
    5 & 5.38 & 13.50 & 33.84 \\
    6 & 5.63 & 13.67 & 35.05 \\
    \bottomrule
  \end{tabular}
\end{table}

At all three scales, a larger $m$ never reduces propagation delay. It is essentially flat across $m$ at $n = 50$ (within 0.4\,s, about 7\% of the mean) and rises gently with $m$ at larger scales: from 11.8 to 13.7\,s at $n = 100$, and from 26.2 to 35.1\,s at $n = 200$. This contradicts the intuition that more neighbors should make propagation faster. The reason is the same trade-off derived in Section~\ref{sec:command-monitoring}: a wider neighborhood carries each command farther per hop, shortening the path (its length scales as $n/m$), but makes each node settle proportionally more keysends at every hop. The two effects cancel, so delay stays $\Theta(n)$ rather than dropping as $\Theta(n/m)$ or compounding as $\Theta(nm)$. The mild residual rise with $m$ reflects the heavier per-keysend settlement that more channels per node add (Section~\ref{sec:scalability}). Botnet size~$n$, not neighbor count~$m$, governs propagation delay.

\subsection{Effect of Injection Point}
\label{sec:injection}

We vary the botmaster's injection position on a fixed D-LNBot chain ($n = 50$, $m = 4$) across five settings: a uniformly random node, the bottom (oldest) end, the middle, the top (youngest) end, and multi-source injection from all three fixed positions simultaneously. Table~\ref{tab:injection} shows the results.

\begin{table}[ht]
  \centering
  \caption{Propagation delay for different botmaster injection positions ($n = 50$, $m = 4$, D-LNBot chain).}
  \label{tab:injection}
  \setlength{\tabcolsep}{12pt}
  \begin{tabular}{@{}l c c@{}}
    \toprule
    \textbf{Injection Position} & \textbf{Delay (s)} & \textbf{Coverage} \\
    \midrule
    Random                           & 7.45 & 100\% \\
    Bottom (oldest node)             & 8.47 & 100\% \\
    Middle                           & 5.68 & 100\% \\
    Top (youngest node)              & 8.48 & 100\% \\
    Multi-source (bottom+middle+top) & 4.18 & 100\% \\
    \bottomrule
  \end{tabular}
\end{table}

Among single injection points the middle is fastest (5.7\,s), roughly 1.5$\times$ faster than either end (about 8.5\,s), because it halves the longest distance a command must travel along the chain. A uniformly random node, averaged over five runs, falls in between (7.5\,s). Multi-source injection from all three positions at once is faster still (4.2\,s): the three simultaneous sources further shorten the longest path, and because forwarding is event-driven, this shorter path outweighs the extra work of handling the duplicate commands. These results are specific to the chain's symmetric structure; other topologies may behave differently.

\subsection{Hardware Resources Consumed by LNTest}
\label{sec:hw-resources}

We measured LNTest's resource consumption by monitoring all Docker containers throughout a run. At $n = 100$ ($m = 4$, D-LNBot topology), the 102 containers (100 C\&C servers, one botmaster, one innocent node) use roughly 8--9\,GB of RAM and average about 4 of the host's 16 CPU cores in steady state, with brief peaks during the channel-opening phase. At $n = 500$, the 502 containers use about 47\,GB of RAM and average roughly 8 cores. Per-node memory grows only slowly with scale (from about 89\,MB to 97\,MB per node), so memory is the binding constraint: a 64\,GB workstation can host on the order of 650--700 nodes, and a 16\,GB laptop around 150. Even at $n = 500$ the CPU stays well below saturation, so these experiments are bounded by propagation delay (Section~\ref{sec:scalability}) rather than by host resources.

\section{Discussion}
\label{sec:discussion}

\subsection{Findings and Implications for Defense}

Our first main finding is that the topology a botnet \emph{actually} forms can differ substantially from the topology its protocol \emph{prescribes}. D-LNBot's formation protocol is designed to produce a uniform chain, but concurrent peer discovery instead yields a \emph{clustered chain} whose bridge nodes are single points of failure (Section~\ref{sec:formation}). This structural weakness does not appear in the published design~\cite{dlnbot} and would be invisible to a defender who reasons only from the protocol specification. The implication is that defenders should profile the \emph{deployed} overlay (for instance, by monitoring channel-open patterns on the blockchain~\cite{ln-centralisation}) rather than assuming the published topology holds.

Second, our takedown experiments reveal that the \emph{type} of overlay topology largely determines which disruption strategy is effective (Section~\ref{sec:takedown}). The D-LNBot chain and the BA scale-free graph are each vulnerable to only one strategy, whereas the autonomous clustered chain fragments under \emph{both}: its bridge nodes are structurally critical and, unlike the D-LNBot chain, distinguishable by degree. The practical takeaway for defenders is clear: a takedown strategy must be matched to the observed topology.

Third, command propagation scales linearly with botnet size ($\Theta(n)$) and is independent of the neighbor count~$m$, not $O(m \log n)$ as D-LNBot claimed~\cite{dlnbot}, because the overlay behaves like a chain rather than a tree (Section~\ref{sec:scalability}). Even at 500 nodes commands still reach every node without partitioning, so LN-based C\&C is practical well beyond proof-of-concept scale. This underscores the urgency of developing detection mechanisms and evaluating them under controlled conditions, which is exactly what LNTest provides. Existing detectors target on-chain Bitcoin botnets~\cite{oneclassclassifier}, whereas LN-based C\&C hides commands in off-chain, encrypted payments.

\subsection{Ethical Considerations}

LNTest is a dual-use tool: the same features that make it valuable for defensive research could be misused to refine real-world LN-based botnets. Similar concerns have long been raised for traditional botnet testbeds and are discussed in dual-use research guidelines for cybersecurity~\cite{dualuse-satori,dualuse-it}. To reduce risk, we deliberately restrict LNTest to Bitcoin \emph{regtest}, bind all Lightning nodes to the host's loopback interface, and focus exclusively on C\&C behavior rather than malware payloads. We recommend that users maintain these constraints, avoid connecting LNTest to public networks, and follow institutional ethics processes before running large-scale experiments.

\subsection{Limitations}

LNTest models only the C\&C layer of an LN-based botnet. We assume that C\&C servers are already compromised machines capable of running full Lightning nodes. In practice, many large botnets rely on compromised IoT devices and home routers with limited resources, as recent campaigns against outdated D-Link devices illustrate\footnote{\url{https://www.bleepingcomputer.com/news/security/malware-botnets-exploit-outdated-d-link-routers-in-recent-attacks/}}. Porting our design to such constrained hardware, or combining LNTest with existing IoT botnet testbeds, is left for future work.

Our experiments use a regtest-based deployment on a single machine, with overlays of up to 500 C\&C nodes. The real Lightning Network has thousands of public nodes, dynamic channel balances, background payment traffic, and non-trivial fee dynamics~\cite{ln-centrality,ln-centralisation}. We also use a single LN implementation (Core Lightning); heterogeneous deployments mixing CLN~\cite{cln}, LND~\cite{lnd}, Eclair~\cite{eclair}, and Rust Lightning~\cite{ldk} may exhibit different routing behavior or failure modes. Furthermore, we assume all C\&C channels have sufficient liquidity and do not model long-lived jamming attacks that could lock HTLC slots and delay command propagation~\cite{congestion-pcn}. Because LNTest also runs entirely on localhost with no real network latency, the propagation delays we report are dominated by Core Lightning's per-hop payment-processing time rather than by network transport, and should be viewed as optimistic lower bounds under idealized conditions.

Finally, our degree-based targeted takedown strategy is deliberately simple: it ranks nodes by degree and removes them in descending order. On the uniform-degree D-LNBot chain this provides no useful signal, which is itself a finding (Section~\ref{sec:takedown}). More sophisticated strategies, such as betweenness-centrality removal~\cite{rohrer2019discharged} or bridge-targeted disruption, may prove more effective. The main value of LNTest lies in comparing \emph{relative} trends across topologies, takedown strategies, and injection points, and in providing a common, reproducible baseline for future work on LN-based botnet defenses.

\section{Conclusion}
\label{sec:conclusion}

We presented LNTest, a self-contained Bitcoin \textit{regtest} testbed for experimentally studying Lightning Network-based botnets. Built from production-grade Bitcoin Core and Core Lightning nodes containerized with Docker, LNTest supports three overlay topology modes (deterministic chain, autonomous peer discovery, and user-supplied graphs) and lets researchers control botnet size, neighbor connectivity, injection point, and takedown strategy as independent variables.

Our experiments produced three main findings. First, D-LNBot's autonomous formation protocol does not produce the uniform chain described in its design; instead, concurrent peer discovery creates a clustered chain with bridge nodes that are single points of failure. Second, command propagation scales linearly with botnet size ($\Theta(n)$) and is unaffected by the neighbor count, not $O(m \log n)$ as previously claimed, because the overlay behaves like a chain rather than a tree. Third, the overlay topology determines which takedown strategy is effective: degree-based targeting devastates scale-free topologies but is useless against uniform-degree chains, random removal shows the opposite pattern, and the autonomous clustered chain, the topology D-LNBot actually produces, is fragile under both strategies. These results demonstrate that defenders must profile the deployed overlay before choosing a disruption strategy.

By releasing LNTest, we provide a reproducible platform for future work on LN-based botnet defenses. Natural extensions include betweenness-centrality-based takedown strategies, heterogeneous Lightning implementations, and integration with IoT botnet testbeds to model resource-constrained deployments.

\begin{credits}
\subsubsection{\discintname}
The authors have no competing interests to declare that are relevant to the content of this article.
\end{credits}

\bibliographystyle{splncs04}
\bibliography{refs}

\end{document}